\documentclass[prl,aps,showpacs,floatfix,byrevtex,amsmath,
superscriptaddress,amssymb,twocolumn]{revtex4}
\usepackage{graphicx}

\begin{document}

\preprint{PREPRINT}
\newcommand{\Cmm}{{C/m$^2$}}
\newcommand{\etal}{{\em et al.\/}}
\title{The effect of entropy on macroions adsorption}

\author{Felipe Jim\'enez-\'Angeles}
\email{fangeles@www.imp.mx}
\affiliation{Programa de Ingenier\'{\i}a Molecular, Instituto Mexicano del Petr\'oleo,
L\'azaro C\'ardenas 152, 07730 M\'exico, D. F., M\'exico}
\affiliation{Departamento de F\'{\i}sica, Universidad Aut\'onoma Metropoloitana-Iztapalapa,
Apartado Postal 55-334, 09340 D.F. M\'exico}

%
\author{Marcelo Lozada-Cassou}
\email{marcelo@www.imp.mx}
\affiliation{Programa de Ingenier\'{\i}a Molecular, Instituto Mexicano del Petr\'oleo,
L\'azaro C\'ardenas 152, 07730 M\'exico, D. F., M\'exico}

\date{\today{}}

\begin{abstract}
We study macroion adsorption on planar surfaces, through a simple
model. The importance of entropy in the interfacial phenomena is
stressed. Our results are in qualitative agreement with available
computer simulations and experimental results on charge reversal
and self-assembling at interfaces.
\end{abstract}

\pacs{68.08.-p, 82.70.Dd, 87.15Aa}

\maketitle

The physics of macroions adsorption at charged interfaces is
subject of recent theoretical and experimental studies
\cite{decher97,rondelez_1998,tanaka_2001,terao_2002,gelbart00,tohver_2001}
because its potential technological applications.
It is known that in the neighborhood of a charged surface,
low-concentrated solutions of monovalent ions
produce an exponentially decaying charge distribution
which is well described by Poisson-Boltzmann equation.
However, adsorption of multivalent ions displays important
deviations from the Poisson-Boltzmann picture and there is not yet
a fully accepted theory which well accounts for this situation
\cite{rondelez_1998,gelbart00}.
In inhomogeneous charged fluids, charge reversal (CR) and charge inversion (CI)
have been reported as mechanisms with which systems minimize free energy.
CR is the overcompensation of the surface's charge by a contiguous
layer of oppositely charged ions from the fluid. Because of
electrostatic equilibrium, next to the CR layer, a second layer of
ions (with same charge sign as that of the surface)  is formed,
producing a CI \cite{attard_1996,kjellander98}.
In molecular engineering, CR and CI are basic mechanisms for
self-assembling polyelectrolyte layers on a charged substrate
\cite{decher97}, reverse mobility experiments \cite{lozada99},
self-assembled DNA-lipid membrane complexes \cite{raedler_sci2},
anomalous macroions adsorption on a Lagmuir film
\cite{rondelez_1998}, and novel colloids stabilization mechanisms
\cite{tohver_2001}.

Although CR was reported since the early 80's
\cite{megen80,mier_1989} and some advances in its understanding
have been done, the mechanism has not been completely understood.
Some authors recognized the relevance of electrostatic and ionic
size correlations in CR \cite{attard_1996,kjellander98,mier_1989}.
In particular, the unsymmetrical size effects (between {\em small}
coions and counterions) has been considered by Greberg and
Kjellander \cite{kjellander98}.
Their analysis of CR and CI is based on mechanistic arguments and
is basically correct, although incomplete: the role of the ionic
concentration, entropy and charge asymmetry is not discussed,
which are not minor effects \cite{tohver_2001}, as we will see.
Recently CR by macroions of an {\em oppositely} charged surface
was considered by Shklovskii {\etal} \cite{shklovskiiL2000}. They
assert in proposing that, when macroions overscreen a charged
surface, these macroions are highly ordered in a bidimensional
Wigner crystal (WC) structure, as a configuration of minimal free
energy.
However, because in the WC theory ionic correlations are only
considered at the level of a two dimensional electrons liquid at
{\em zero} temperature, this approach leads to incorrect
conclusions. For example, it is predicted that CR decreases by
increasing surface charge density (see Fig. 3 in
ref.~\cite{shklovskiiL2000}) whereas molecular simulations show
the opposite \cite{jimenez_2001,terao_2001}.
Nevertheless, WC theory has a good agreement with non-zero Monte
Carlo simulations of a two-dimensional electron liquid
\cite{totsuji_1978} and zero temperature molecular dynamics
calculations of a two-dimensional layer of adsorbed ions
\cite{messina00}.
The role of entropy in the attraction of like-charged particles
has been recognized in the past \cite{muthuJCP_1996}. Here we will
analyze the effect of electrostatic and short range correlations
on macroions adsorption at interfaces.  A decreasing of the system
entropy always produces an increasing of macroions adsorption.
Thus, it is concluded that self-assembling at interfaces, CR, CI
and like-charged attraction are {\em mainly} ruled by entropy.

For the present study we consider a uniformly charged planar
surface with $\sigma_0$ charges per unit area contiguous to a
macroions solution. In our model, we include the two relevant
forces of the system: the long range coulombic interaction and the
hard-core interaction forces. Therefore macroions are considered
to be charged hard spheres (with charge $Q=Ze$, diameter $a$ and
number density $\rho$; being $e$ the electronic charge) plus an
additional amount ($\rho_c$) of neutralizing counterions (with
charge $q=z_c e$ and radius $r_0=a_c/2$). The mixture is embedded
in a continuous media of dielectric constant $\epsilon$, set equal
to that of the charged surface in order to avoid image charges. By
the electroneutrality condition
$-\sigma_0=\int_{0}^{\infty}\rho_{el}(x)dx$ with $\rho_{el}=\sum_{i=1}^{2}q_i\rho_i(x)$,
and $\rho_i(x)$ being the local concentration profile of the
$i$-th species. Our model is an extension to that considered by
Kjellander and Greberg \cite{kjellander98}. The WC theory
\cite{shklovskiiL2000} has some similarity with our model.
However, whereas we use direct (macroion-macroion,
macroion-counterion and macroion-wall) coulombic interactions, in
WC screened interactions are used. More importantly,
short range correlations and entropy (of
fundamental importance for finite temperature calculations) are
included in our theory \cite{jimenez_2001}.

We study the system by means of integral equations using the
hybrid closure {\em hypernetted chain/mean spherical
approximation} (HNC/MSA). This approximation, for simpler models
of charged inhomogeneous fluids, has shown to be in agreement with
computer simulations where CR and other effects are studied
\cite{jimenez_2001,lozadaPRL_1997}. For the present model, we
found qualitative agreement with integral equations results
\cite{kjellander98} and molecular simulations \cite{terao_2002}.
The HNC/MSA equations can be derived in an straightforward manner
\cite{henderson92a} and are given by
\begin{equation}
g_i(x)=\exp \left\{-\beta u_{\alpha i}(x)+ \sum_{j=1}^{2} \rho_j
\int h_{j}(y)c_{ij}(s)dv_{3}\right\}, \label{hnc:msa}
\end{equation}
where $\rho_i$ is the bulk concentration of the $i$-th species,
$h_{j}(x) \equiv g_{j}(x)-1$ and $c_{ij}(s)$ are the total and
direct correlation functions, respectively; $u_{\alpha i}(x)$ is
the ion-surface interaction potential, $\beta = 1/k_{\rm B}T$ with
$k_B$ the Boltzmann constant and $T$ the absolute temperature.
\begin{figure}
\includegraphics[width=7.0cm]{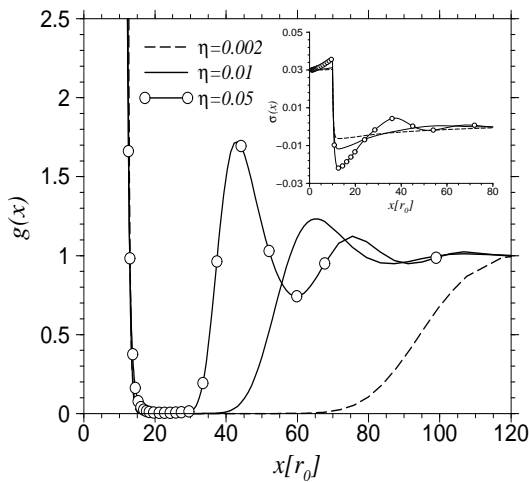}
\caption{Reduced density profile $g(x)$ for negative macroions
($Z=20$) next to an positively charged surface
($\sigma_0=0.03$\Cmm). We present results for $\eta=0.002,0.01,
0.05$. In the inset we show the corresponding adsorbed charge
densities $\sigma(x)$.} \label{FIG1+}
\end{figure}
From HNC/MSA, the concentration profiles (${\rho_i(x)=\rho_i g_i
(x)}$) are obtained directly. These are related to the ion-surface
potential of mean force, $w_i(x)$, by ${g_i(x)= \exp \{-\beta
w_i(x)\}}$. Eq. \ref{hnc:msa} can be written as $w_{i}(x) = -k_B T
\ln[g_{i}(x)]= \psi_{i}(x) + J_i(x)$, where $\psi_{i}(x)$ is the
mean electrostatic potential and $J_{i}(x)$ are the short-range
contributions. Since both, $\psi(x)$ and $J_{i} (x)$ are
functionals of $\rho _{el} (x)$ and are in a non-linear equation,
the charge and size correlations are {\em non-independent}. In our
theory macroions adsorption is related to both: charge and ionic
size correlations. In the Poisson-Boltzmann description the short
range correlations are not considered in any way (i. e.,
$J_i(x)=0$). Therefore, PB does not describe correctly many of the
adsorption phenomena.

In a system at constant volume V, fixed particle number
N and T, its entropy is given by
$S \equiv S^{\rm id}+ k \left(\frac{\partial \ln Z_N}{\partial
T}\right)_{N,V}$,
where $S^{\rm id}$ is the ideal gas entropy and the
configurational integral $Z_N=\int...\int\exp\left\{-\beta
U_N\right \}d{\bf r}_1,...,d{\bf r}_N$, being $U_N$ the potential
energy of the N interacting particles. The exact computation of
$S$ is not an easy task. However, from the expressions for $s$ and
$Z_N$ is clear that: (i) the accessible volume \cite{my_note}
decreases as the volume fraction
($\eta_T\equiv\frac{\pi}{6}\sum_{i}\rho_i a_i^3$) increases. Thus,
the larger $\eta_T$ the lower the entropy. (ii) Electrostatic
interactions between like-charged particles also reduce the system
number of accessible configurations (hence entropy) by increasing
the strength of the coulombic coupling interaction $\xi=\beta
q_{i}^2/\epsilon a_{i}$. Less entropy {\em always} implies higher
order, which in an inhomogeneous fluid translates in higher
adsorption, as we will show. In this picture it is plausible
charge reversal by monovalent ions, macroions adsorption on
uncharged (or like-charged) surfaces, as well as the {\em strong}
dependence of CR on $\eta$. These effects can not be explained
only with energy arguments.

\begin{figure}
\includegraphics[width=6.5cm]{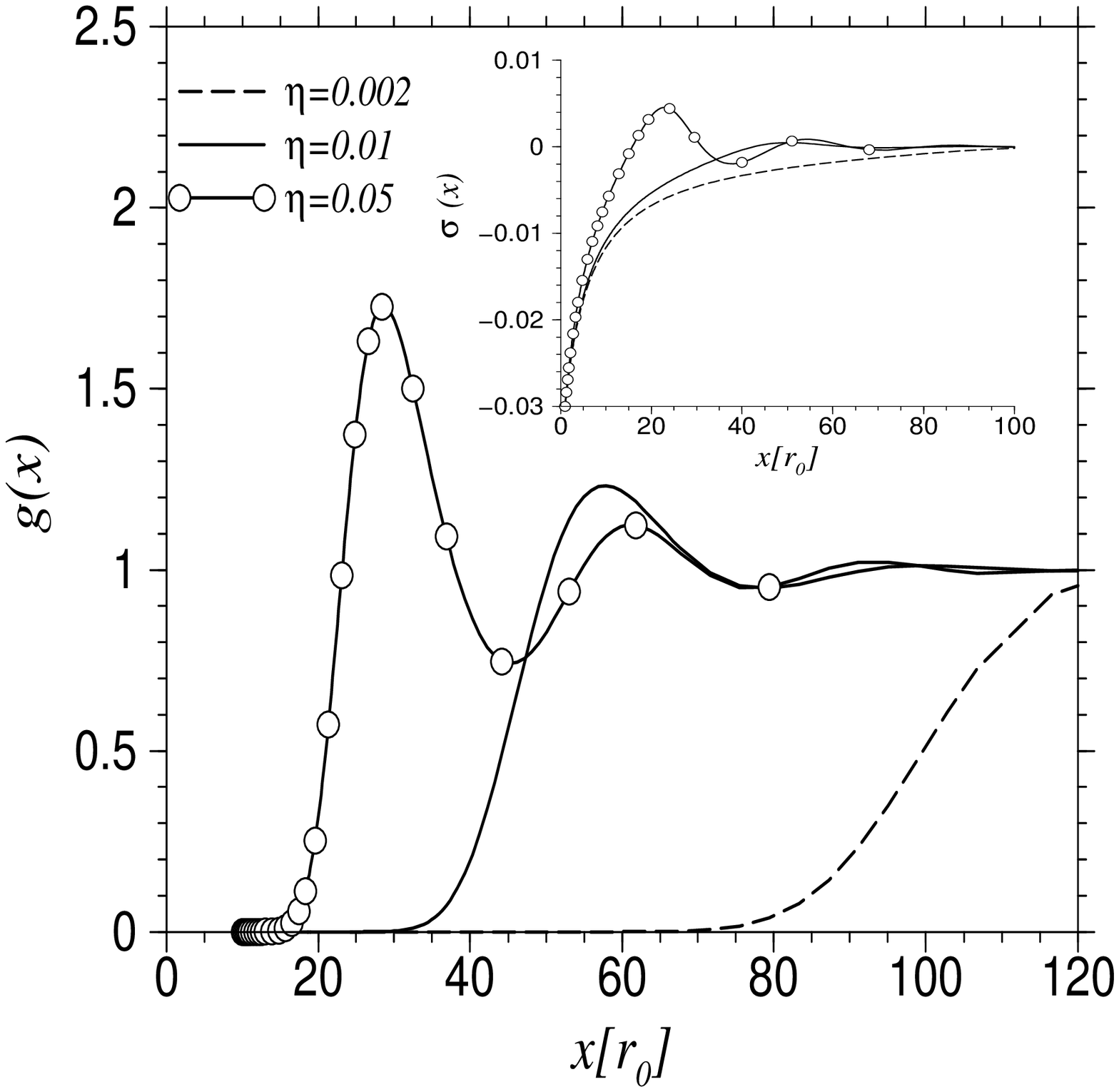}
\caption{Same as Fig.~\ref{FIG1+} but $\sigma_0=-0.03$\Cmm.}
\label{FIG1}
\end{figure}

We will discuss our results in terms of $g_{i}(x)$, the  local net
charge per unit area $\sigma (x) = \sigma_0+{\int\limits_{0}^{x}
{\rho _{el} (y)dy}}$ and the absolute maximum of this function
($\sigma^{*}\equiv {\rm max}\{\sigma(x):x\in[0,\infty)\}$);
with the charge profile given by
$\rho _{el} (x) \equiv {\sum\limits_{m = 1}^{2} {ez_{m}} } \rho _{m} g_{m} (x)$.
The effective electrostatic force on an ion is $f_{i}^{e} (x)
\equiv \frac{2\pi}{\varepsilon}q_i\sigma (x)$. Therefore, $\sigma
(x)$ not only quantifies charge adsorption but also the
wall-particle effective electrical force. We have solved HNC/MSA
for several values of $Z$ and $\rho$. In all cases the little ions
(counterions) are monovalent,
 i.e., $|z_c|=1$. In
all our calculations we have used fixed values of $a=4$nm,
$a_c=0.4$nm, $T=298$K and $\epsilon=78.5$. We present results for
macroions next to positively charged, negatively charged and
uncharged surfaces. We study the influence on macroions adsorption
of $\sigma_0$, $Z$, $\eta$.

In Fig.~\ref{FIG1+} we show the macroions reduced concentration
profile (RCP) for {\em negative} macroions ($Z=20$, i.e.,
$Q=20e<0$) next to a positively charged surface
($\sigma_0=0.03$\Cmm) for three different macroions concentrations
($\eta=0.002,0.01,0.05$).
The macroion RCP shows a very strong adsorption to the wall and a
considerable amount of counterions surrounding
macroions: In all figures, the counterions RCPs are not plotted
for clarity. A second layer of macroions is adsorbed mediated by
counterions.
In the inset we show the corresponding $\sigma (x)$ functions: At
a distance of one macroion radius a deep minimum is observed,
corresponding to a very strong CR. For $\eta=0.05$, the maximum
located around $x=35r_0$ shows a CI. In the inset, the maximum
located at $x=a/2$ indicates that small positive ions are adsorbed
to the wall. Hence, positive charge is accumulated next to the
positively charged surface. We denominate this effect as {\em
overcharging}.
This effect is energetically unfavorable,
thus, it must be attributed to the short range correlations
since it increases with $\eta$.
This prediction agrees qualitatively with computer simulation of
Tanaka and Grosberg \cite{tanaka_2001}. The effective wall
electrical field, which is proportional to $\sigma (x)$, is
positive before the first layer of macroions and then negative,
before the second layer.
Therefore, the electrical force is first attractive and then
repulsive to negative ions. The behavior of the total force on an
ion of species $i$, however, is implicit in the RCP, i.e., a
$g_{i} (x)$ above (below) 1 implies that $F_{i} (x)$ is attractive
(repulsive).

Some of the physics of macroions adsorption is not explicit when
considering macroions adsorbed on oppositely charged surfaces.
In Fig.~\ref{FIG1} we show the macroions RCP for negative
macroions ($Z=20$) next to a negatively charged surface
($\sigma_0=-0.03$\Cmm).
Counterions adsorb on the charged surface (not shown) and, for
sufficiently high values of $\eta_T$, a layer of macroions is
adsorbed to the wall, after a relatively large {\em void} of
macroions. Hence a surface CR is induced by little ions (according
to the $\sigma(x)$ functions in the inset) and they, in turn,
induce macroions {\em adsorption} and the formation of the
macroions layer. For sufficiently low values of
$\eta=\frac{\pi}{6}\rho a^3$ there is no macroions adsorption.
The qualitative agreement of our Fig.~\ref{FIG1} results for
$\eta=0.01$, with Monte Carlo data reported in
ref.~\cite{terao_2002}, is encouraging. From Figs.~\ref{FIG1+} and
~\ref{FIG1} it is seen that higher $\eta_T$ produces higher
adsorption. This prediction is in agreement with computer
simulations ~\cite{tanaka_2001,terao_2001,jimenez_2001}.

\begin{figure}
\includegraphics[width=7.0cm]{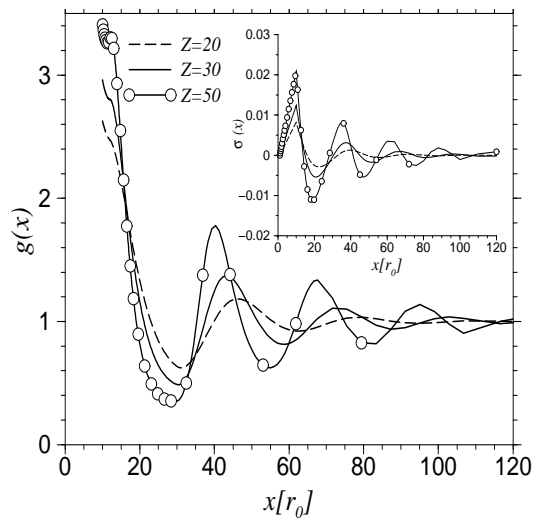}
\caption{Reduced density profile $g(x)$ for macroions next to a
uncharged surface. We present results for $Z=20, 30, 50$;
$\eta=0.05$. In the inset are the corresponding adsorbed charge
densities $\sigma(x)$.} \label{FIG2}
\end{figure}

In Fig.~\ref{FIG2} the RCP are shown for negative macroions
next to an {\em uncharged} surface. For this
macroions volume fraction ($\eta=0.05$) we observe colloids adsorption
even for low charged colloids [as low as $Z=1$ (not shown)].
Since the wall is uncharged, the strong macroions adsorption and
wall overcharging is due to the low system entropy, which in this
case is due to both: high $\eta$ and high $\xi$. Increasing $Z$
increases ($\xi$), and thus decreases the entropy and increases
macroions adsorption and inducing a surface charge. This induced
charge is in agreement with the nanoparticles halo reported by
Tohver~\cite{tohver_2001}, which seems to be a new colloidal
stabilization mechanism.  Our results show that there is not a
minimum surface or macroions charge for the occurrence of CR.
Therefore, the assembled nanostructures at interfaces are
stabilized by entropy-induced attractive forces, rather than the
surface-macroions electrostatic correlations as it is proposed in
the the WC theory. These findings are in agreement with
experiments on polyelectrolyte adsorption where it is found that
CR of a charged substrate ``is more a property of the polymer than
a property of the surface'' \cite{decher97}, and with previous
theoretical results, where the importance of particle's excluded
volume (entropy) is recognized as an attractive force mechanism
between like charged particles \cite{muthuJCP_1996}. Please notice
that CR in Fig.~\ref{FIG1} through Fig.~\ref{FIG2}, is of the same
order of magnitude whether the substrate charge is positive,
negative or zero.

\begin{figure}
\includegraphics[width=7.0cm]{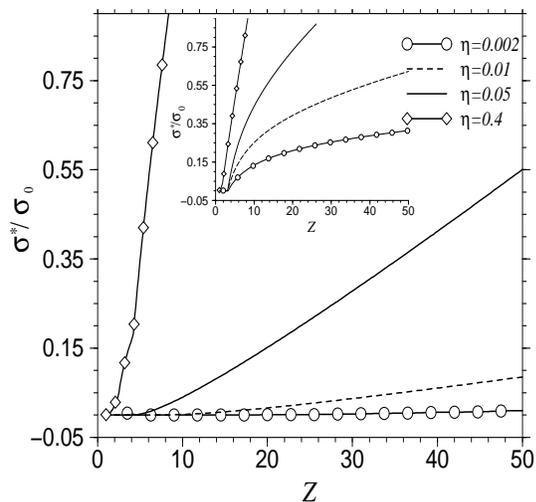}
\caption{$\sigma^*/\sigma_0$ as a function of the, negative,
macroions valence $Z$ with $\sigma_0=-0.03$ , for $\eta=0.002,
0.01, 0.05, 0.4 $. In the inset we show the results when
$\sigma_0=0.03$\Cmm.} \label{TAPE27s}
\end{figure}

In Fig.~\ref{TAPE27s}, $\sigma^*/\sigma_{0}$ is plotted as a
function of $Z$ for $\eta=0.002,0.01,0.05$, for both
$\sigma_0=-0.03$\Cmm and $\sigma_0=0.03$\Cmm (in the inset). For
$\sigma_0=-0.03$\Cmm, macroions and surface are both negatively
charged, and CR is carried out by the small counterions (see
discussion of Fig.~\ref{FIG1}). Nevertheless $\sigma^*/\sigma_{0}$
strongly depends on $\eta$, being negligible for low $\eta$. For
$\sigma_0<0$, $\sigma^*/\sigma_{0}$ monotonically increases as $Z$
with a positive concavity
(${\frac{1}{\sigma_0}\frac{d^2\sigma^*}{d^2Z}>0}$) whereas for
$\sigma_0>0$ ${\frac{1}{\sigma_0}\frac{d^2\sigma^*}{d^2Z}<0}$.
These results imply that oppositely charged macroions are much
more efficiently adsorbed into a charged surface.
Our results for oppositely-charged macroions resemble those shown
in Fig.~{3} of ref.~\cite{shklovskiiL2000} (WC theory) where
$\sigma^*/\sigma_{0}$ increases in the same way with $Z$ and
${\frac{1}{\sigma_0}\frac{d^2\sigma^*}{d^2Z}>0}$.
However, the role of entropy is not properly considered in the WC
theory~\cite{shklovskiiL2000}: in WC there is a minimum in
$\sigma_{0}$ and $Z$ must be large to have overcharging which is
in opposition to computer simulations \cite{jimenez_2001} and our
theory. When entropy is considered we observe the following:
i) $\sigma^*/\sigma_{0}$  depends more strongly on $\eta$ than on
$Z$ \cite{my_note2}.
ii) For a sufficiently high value of $\eta$ our theory predicts
$\sigma^*>0$, even for $Z=1$, independently of the value of
$\sigma_0$ (even for $\sigma_0=0$, see Fig. \ref{TAPE27s}).
iii) Although not shown explicitly, we found that
$\sigma^*/\sigma_{0}$ (i.e., CR) increases by increasing
$|\sigma_0|$ (positive or negative, not shown), which is in
agreement with recent simulations results
\cite{jimenez_2001,terao_2001} and, as we pointed out above, in
opposition to the WC theory.

From Figs. \ref{FIG1} to \ref{TAPE27s} it is clear that macroions
adsorption strongly depends on $\eta_T$. This effect is far more
important for CR than surface or ionic charge. In an experiment of
macroion adsorption on a Langmuir film of amphiphilic molecules
Rondelez \cite{rondelez_1998} finds that macroions adsorption is
increased (hence CR increases) as the lipid layer charge density
decreases, by increasing the surface area. This is consistent with
our results since an increase of the area per molecule decreases
the surface charge density but not the macroions adsorption which
is strongly driven by macroions excluded volume.  According to
this the CR should also increase by increasing the macroions size
and/or concentration. In another important experiment R\"adler
{\etal} find that the effective charge on DNA-cationic liposome
complexes depends on the DNA-to-cationic lipid mixing ratio, i.e.,
for a certain DNA-to-DNA distance, $d_{DNA}$=$d_{iso}$ the complex
is neutral. However, if the DNA solution concentration is
increased $d_{DNA}$ becomes lower than $d_{iso}$, which implies
higher DNA adsorption on the membranes, i.e., CR on the membrane.
This effect can not be explained by energy arguments, but
according to our findings an increase in macroion concentration
produces higher adsorption.

In summary, a model for macroions solution next a charged surface
has been considered.
We found that entropy plays a fundamental role in macroions
adsorption phenomena and, in consequence, for assembly of
organized nanostructures at interfaces. Some of our predictions
seem to be in agreement with previous theoretical studies of
polyelectrolyte solutions \cite{muthuJCP_1996}, computer
simulations \cite{jimenez_2001,tanaka_2001,terao_2002} and
experimental results
\cite{decher97,rondelez_1998,tohver_2001,raedler_sci2}.

We gratefully acknowledge the financial support of INDUSTRIAS NEGROMEX.


\begin{thebibliography}{10}
\bibitem{my_note}
The accessible volume is defined as the free unoccupied volume
left by the
  fluid particles.
\bibitem{my_note2}
While an increment of $\eta$ (say by a given factor) strongly
increases
  $\sigma^*/\sigma_{0}$, an increment of $Z$ (by the same factor), does not
  produce a comparable increment.
\end{thebibliography}

\end{document}